\documentclass[preprint]{elsarticle}
\usepackage{lineno,hyperref,exscale}
\usepackage{graphics,color}
\usepackage[normalem]{ulem}
\modulolinenumbers[5]

\journal{Journal of Magnetism and Magnetic Materials}

\newcommand{\teff}{T$_{\mbox{\tiny eff~}}$}
\newcommand{\be}{\begin{eqnarray}}
\newcommand{\ee}{\end{eqnarray}}
\global\long\def\av#1{\left\langle #1 \right\rangle}

\begin{document}

\begin{frontmatter}

\title{Non-linear quantum critical dynamics and fluctuation-dissipation ratios far from equilibrium}

\author{Farzaneh Zamani}
\address{Max Planck Institute for the Physics of Complex Systems, N\"othnitzer Str. 38, 01187 Dresden, Germany}
\address{Max Planck Institute for Chemical Physics of Solids, N\"othnitzer Str. 40, 01187 Dresden, Germany}
\author{Pedro Ribeiro}
\address{CeFEMA, Instituto Superior Técnico, Universidade de Lisboa, Av. Rovisco Pais, 1049-001 Lisboa, Portugal}
\address{Russian Quantum Center, Novaya Street 100 A, Skolkovo, Moscow Area, 143025 Russia}
\author{Stefan Kirchner}
\address{Center for Correlated Matter, Zhejiang University, Hangzhou,  Zhejiang 310058, China}
\ead{stefan.kirchner@correlated-matter.com}

\begin{abstract}
Non-thermal correlations of strongly correlated electron systems and the far-from-equilibrium properties of phases of condensed matter have become a topical research area.  
Here, an overview of the non-linear dynamics found near continuous zero-temperature phase transitions within the context of {\it effective} temperatures is presented. 
In particular, we focus on models of critical Kondo destruction. Such a quantum critical state, where Kondo screening is destroyed in a critical fashion, is realized in a number of rare earth intermetallics. This  raises the possibility of  experimentally testing for the existence of fluctuation-dissipation relations far from equilibrium in terms of  effective temperatures near quantum criticality. 
Finally, we  present an analysis of a non-interacting, critical  reference system, the pseudogap resonant level model, in terms of effective temperatures and contrast these results with those obtained near interacting quantum critical points.
\end{abstract}


\end{frontmatter}


\section{Introduction}
The unusual properties observed in a growing number of strongly correlated metals that are close to a continuous zero temperature phase transitions   at the brink of magnetism have resulted in
attempts of addressing the critical properties within a description beyond the Ginzburg-Landau-Wilson paradigm. In the context of heavy fermion compounds, where the existence of unconventional quantum criticality has been experimentally verified, it has become clear that 
one of the key questions concerns the fate of the Kondo effect as the lattice undergoes a magnetic transition as a function of a tuning parameter, {\it e.g.} pressure or magnetic field, and at zero temperature~\cite{Si2001,Coleman.01}. A particularly promising alternative to the traditional Hertz-Millis-Moriya or spin-density wave (SDW) scenario~\cite{Hertz.76,Millis.93} is local quantum criticality (LQC) where Kondo screening is critically destroyed at the quantum critical point (QCP)~\cite{Si2001,Si.14}.
Within LQC, the QCP is characterized by a scale-invariant spectrum that results in dynamical or $\omega/T$-scaling of correlation functions. This is in line with a growing number of experimental results~\cite{Si.14}.

The study of correlated matter, including the heavy fermion systems, has undergone a considerable evolution over the past decades  that led from the measurements of  thermodynamic quantities like {\it e.g.} specific heat to modern  microscopic probes of dynamic correlation functions.   
Even for heavy fermion compounds which are dominated by an enhanced density of states in close vicinity of the Fermi energy and concomitantly small characteristic
energy scales, these spectroscopic probes, {\it e.g.} angle-resolved photoemission spectroscopy (ARPES)~\cite{Koitzsch.09,Xiao-Wen.11} and scanning tunneling spectroscopy (STM)~\cite{Aynajian.10,Schmidt.10,Ernst.11} have recently become available. Thus, ARPES, STM, inelastic neutron scattering, etc. are capable  of determining dynamic correlation functions and their scaling properties near quantum criticality. The fluctuation-dissipation theorem (FDT), valid in the linear response regime, connects the measured response functions to the correlation functions of the system. For a  quantum critical system, possessing a scale-invariant spectrum, {\it i.e.} in the absence of any intrinsic scale, any perturbation may probe the system in the non-linear regime beyond the validity of the FDT~\cite{Kirchner.09}.
Understanding the properties of quantum criticality out of equilibrium has accordingly become  a topical issue. This applies to both its theoretical description as well as its experimental signatures.

Here, we address the concept of  effective temperatures  in the context of quantum critical steady states far from equilibrium, both for the SDW as well as for a model of critical Kondo destruction, and present an analysis of a non-interacting, critical  reference system, the pseudogap resonant level model, in terms of effective temperatures and contrast these results with those obtained near fully interacting quantum critical points.
  
\section{Effective temperatures near quantum critical steady states}
\label{Sec:QCP}
The notion of an effective temperature (\teff) has  originally been introduced in the context of turbulence by Hohenberg and Shraiman~\cite{Hohenberg1989} and later extended to
glassy systems displaying slow, relaxational dynamics~\cite{Cugliandolo.97}. The underlying idea is that specific features of an out-of-equilibrium problem can be described in terms of a thermal ensemble at an effective temperature \teff. A formal definition of \teff
is based on the fluctuation-dissipation theorem (FDT) which at the classical level states that  the correlation ($C$) and response ($R$) function are related by
\begin{equation}
\label{Eq:classicalFDT}
TR(t-t')=\frac{\partial}{\partial t} C(t-t').
\end{equation}
The realm of validity of the FDT is confined to the linear-response regime but Eq.~(\ref{Eq:classicalFDT})  can be used to introduce an \teff in the non-linear regime, as demonstrated in Ref.~\cite{Cugliandolo.97}.
For a recent review on the concept of effective temperatures in classical and quantum statistical mechanics, see {\it e.g.} Ref.~\cite{Cugliandolo.11}.

In general, the \teff defined through Eq.~(\ref{Eq:classicalFDT}) will depend on the correlation function under consideration which might limit the significance of such a quantity in the description of the out-of-equilibrium dynamics.
In the context of classical critical systems, {\it i.e.} systems close to a finite-temperature continuous phase transition, the usefulness and range of validity of the notion of \teff has been explored in a number of works~\cite{Godreche.00,Calabrese.04,Corberi.07,Kroha.07,Cugliandolo.11}.
In particular, Calabrese and Gambassi were able to show that even if \teff  is observable independent at the level of the Gaussian or mean field approximation, the inclusion of
higher order contributions beyond the Gaussian level will in general make \teff observable dependent~\cite{Calabrese.04}.

In what follows,  we will restrict ourselves to the  simplest possible case, {\it i.e.}, to purely dissipative dynamics, which is often referred to as model A dynamics~\cite{Hohenberg.77}: Within the Ginzburg-Landau-Wilson framework, the time derivative of the order parameter is related to the order parameter derivative of the free energy functional. Depending on whether the order parameter is conserved or not, the relaxation rate in the long-wavelength limit does or does not vanish.

The quantum version of the FDT states that for a bosonic correlator
\begin{equation}
\label{Eq:QFDT}
R(\omega,T)=\Big[\tanh\big(\frac{\omega}{2T}\big)\Big]C(\omega,T).
\end{equation}
Within the non-equilibrium Green function formalism on the Keldysh contour, the FDT reflects itself in the relation between the larger ($G^>$) and lesser ($G^<$) and the retarded ($G^R$) and advanced ($G^A$) (fermionic) Green functions~\cite{Ribeiro.15}
\begin{equation}
\label{Eq:GF}
G^>(\omega,T)+G^<(\omega,T)=\Big[\tanh\big(\frac{\omega-\mu}{2T}\big)\Big]\Big(G^R(\omega,T)-G^A(\omega,T)\Big).
\end{equation}
The combination $G^>(\omega,T)+G^<(\omega,T)$ is often called the Keldysh Green function.
We will refer to $FDR_G=(G^>(\omega,T)+G^<(\omega,T))/(G^R(\omega,T)-G^A(\omega,T))$ as the fluctuation-dissipation ratio of quantity $G$.
In the high-temperature limit, the analog of equation Eq.~(\ref{Eq:QFDT}) for a bosonic correlator reduces to the classical form, Eq.~(\ref{Eq:classicalFDT}).
 
\section{Critical Kondo destruction and SDW: quantum critical relaxational dynamics}

In this section, we will discuss the  possible existence of \teff for current-carrying steady states near the SDW QCP and the critical Kondo destruction QCP. 

\subsection{\teff for the SDW QCP}
The field  theoretic description of the QCP of the SDW type  is based on the action~\cite{Hertz.76}
\begin{eqnarray}
\label{Eq:QuantumSDW}
S &=& S_2+S_4\,=\,T \sum_{\omega_n}\int d\vec{q} \Big(\delta +q^2 +\frac{|\omega_n|}{\gamma q^a}\Big)\big |\Phi(\vec{q},\omega_n)\big |^2\nonumber \\
&+&\frac{u}{4!}\delta(\vec{q}_1+\vec{q}_2+\vec{q}_3+\vec{q}_4)\,\delta_{n_1+n_2+n_3+n_4,0}\,\prod_{i=1,4}\,\sum_{\omega_{n_i}}\,\int d\vec{q}_i\Phi(\vec{q}_i,\omega_{n_i}) ,
\end{eqnarray}
where $\omega_{n_{i}}$ are Matsubara frequencies, the $\vec{q}$ labels momentum, $\delta$ is the mass gap that vanishes in a powerlaw fashion at the QCP, and  $a=0 ~(a=1)$ for an anti-ferromagnet (ferromagnet).
The Gaussian part, $S_2$ of Eq.~(\ref{Eq:QuantumSDW}), generalized onto the Keldysh contour becomes
\begin{eqnarray}
\label{Eq:GaussianKeldysh}
S_2=-i \int_{-\infty}^{\infty}dt \,dt' \,( \Phi_{\mbox{\tiny cl}}^{*}(\vec{q},t)\,\,\Phi_{\mbox{\tiny qm}}^{*}(\vec{q},t))\left( \begin{array}{cc}
0 & (\chi^{-1})^A  \\
(\chi^{-1})^R & (\chi^{-1})^K \end{array} \right)\left( \begin{array}{c}
\Phi_{\mbox{\tiny cl}}^{}(\vec{q},t')  \\
\Phi_{\mbox{\tiny qm}}^{}(\vec{q},t')  \end{array} \right),
\end{eqnarray}
where the fields $\Phi_{\mbox{\tiny cl}}$ and $\Phi_{\mbox{\tiny qm}}$ are related to the fields on the forward ($\Phi^+_{\mbox{\tiny }}$) and backward ($\Phi^-_{\mbox{\tiny }}$) component on the Keldysh contour via a canonical transformation, see {\it e.g.} Ref.~\cite{Kamenev2009}. The FDR of the order parameter susceptibility at the Gaussian level can be obtained  from $\chi^K$ and $\chi^R-\chi^A$ of Eq.~(\ref{Eq:GaussianKeldysh}). 

The generalization of the SDW action, Eq.~(\ref{Eq:QuantumSDW}), onto the Keldysh contour, necessary for a  proper description  of 
non-thermal steady states has been  {\it e.g.} considered in Refs.~\cite{Mitra.06,Mitra.08,Hogan.08}. 
The authors of Ref.~\cite{Mitra.06} considered the current-carrying steady-state of a two-dimensional ferromagnetic film in proximity to its QCP, where $T_{\mbox{\tiny Curie}} \rightarrow 0$ and made the insightful observation that this system is placed at its upper critical dimension as the order parameter is no longer conserved.
The non-linear conductance of such a setup, {\it i.e.} a thin films of CaRuO$_3$, which  may be located in close proximity of a ferromagnetic QCP, has been reported in Ref.~\cite{Esser.14}.
In the static, long-wavelength limit, the spin susceptibility  is found to obey~\cite{Mitra.06}
\begin{equation}
\label{ChiAdv}
(\chi^{-1})^A=\delta+\xi q^2+i\frac{\omega}{\gamma},
\end{equation}
\begin{equation}
\label{ChiKeldysh}
(\chi^{-1})^K=-2i\sum_{\alpha,\alpha'=L,R} \tilde{V}_{\alpha} \tilde{V}_{\alpha'} \coth\Big(\frac{\omega+\mu_\alpha-\mu_{\alpha'}}{2T}\Big)\frac{\omega+\mu_\alpha-\mu_{\alpha'}}{\gamma},
\end{equation}
where $\gamma$ is the damping coefficient and $V=(\mu_L-\mu_R)/e$ is the applied bias voltage, maintaining the steady state and $\tilde{V}_L$ ($\tilde{V}_R$) is the coupling strength between ferromagnetic film and the left (right) lead.

The perturbative RG procedure of Ref.~\cite{Mitra.06} dealing with the $S_4$ analog on the Keldysh contour preserves the structure of Eq.~(\ref{Eq:GaussianKeldysh}) except for a possible change in the prefactor of Eq.~(\ref{ChiKeldysh}). 
From Eq.~(\ref{ChiKeldysh}) one finds in the large-$V$ limit~\cite{Mitra.06} 
\begin{equation}
\chi^K=\frac{(\chi^{-1})^K}{(\chi^{-1})^A(\chi^{-1})^R}=\frac{4iV_L V_R V/\gamma}{(\delta+\xi q^2)^2+(\omega/\gamma)^2}.
\end{equation}
Thus, one is lead to expect in the large-$V$ limit that \teff is proportional to $V$:
\begin{equation}
\label{Eq:TEFF}
T_{\mbox{\tiny eff}}=4\tilde{V}_L\tilde{V}_R\,V.
\end{equation}

Note that the nonequilibrium distribution function of $\chi^{-1}$ is a direct consequence of the distribution function of the one-particle Green function of the electrons in the film which is a
double step function due to the coupling to the left and right lead which are kept at chemical potential $\mu_L$ and $\mu_R$ respectively. This is reminiscent of what happens in the resonant level model, see Sec.~\ref{Sec:RLM}.\\[1.2ex]

\subsection{\teff for critical Kondo destruction}
\begin{figure}[b!]
\centering{}\includegraphics[width=0.75\linewidth]{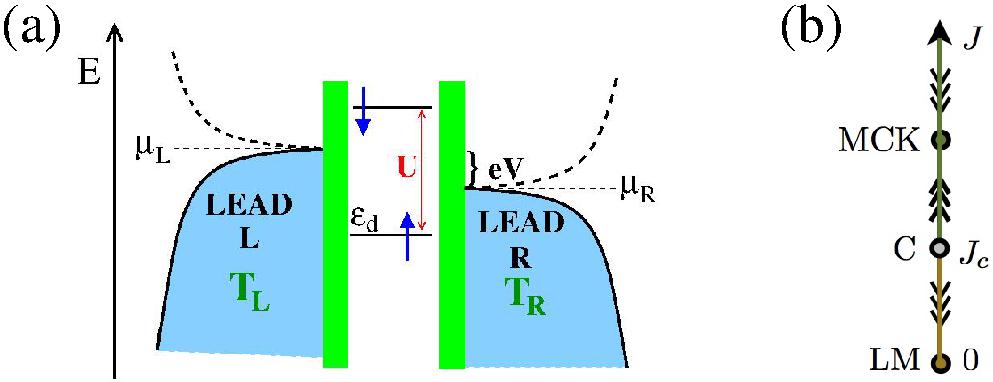}
\protect\caption{\label{fig:dotsetup}
{\bf Schematic  of the pseudogap Anderson model and flow diagram of the pseudogap Kondo model} {\bf (a)} A local level, characterized by the energy $\varepsilon_d$  of the singly occupied state and the Coulomb repulsion $U$ for the doubly occupied level
is hybridizing with two identical electron bands  possessing density of states that vanish in a powerlaw fashion at their respective Fermi levels. Energies are measured from the chemical potential $\mu=\mu_L=\mu_R=0$. The bias voltage  $V$ is given by $V=(\mu_L-\mu_R)/e$. In the limit where $U,\varepsilon \ll \Gamma$, where $\Gamma$ is the level broadening due to the hybridization, the system can be mapped via a Schrieffer-Wolff transformation to the pseudogap Kondo model, whose dynamical large-N limit is considered in Sec.\ref{Sec:QCP}. If, on the other hand, $U=0$, the model turns into the pseudogap resonant level model discussed in Sec.\ref{Sec:RLM} .
{\bf (b)} Flow diagram of the SU(N)$\times$SU(M) pseudogap Kondo model. A local quantum critical point (C) separates a phase with Kondo screening, governed  by a critical fixed point (MCK) from the weak coupling local moment phase (LM), provided the powerlaw exponent $r$ obeys $0<r<r_{\mbox{\tiny max}}$.
}
\end{figure}
We now turn to a discussion of \teff in the context  of critical Kondo destruction~\cite{kirchner2010,Ribeiro2013b,Ribeiro.15}. The results discussed below are based on the dynamical large-N limit of the pseudogap Kondo model. N here is related to the spin symmetry group SU(N) of the local degree of freedom which has been enlarged  from the original SU(2). The dynamical large-N limit on the Keldysh contour has been discussed in detail in Refs.~\cite{Kirchner.09,Ribeiro2013b}. This approach has the great advantage that it treats thermal and non-thermal steady state on equal footing and thus allows for an unbiased comparison between the equilibrium and out-of-equilibrium behavior. 

The pseudogap Kondo model~\cite{Withoff1990} describes a quantum spin that is  coupled anti-ferromagnetically  to itinerant electrons of half-bandwidth $D$ possessing a density of states that vanishes in a powerlaw fashion at the Fermi energy ($\omega=0$), $\rho(\omega)\sim |\omega|^r\Theta(D-|\omega|)$ with power exponent $0<r<r_{\mbox{\tiny max}}<1$. 
The SU(N)$\times$S(M) symmetric version of the model is given 
by the Hamiltonian
\begin{eqnarray}
H & = & \sum_{p,\alpha\sigma l}\varepsilon_{pl}c_{p\alpha\sigma l}^{\dagger}c_{p\alpha\sigma l}+\frac{1}{N}\sum_{ll'}\sum_{\alpha}J_{ll'}\mathbf{S}.\mathbf{s}_{\alpha;ll'},
\label{eq:Hamiltonian_PGK}
\end{eqnarray}
where $J_{ll'}>0$ and $\sigma=1,\ldots,N$ and $\alpha=1,\ldots,M$ are, respectively, the $SU(N)$-spin
and $SU(M)$-channel indices, $l=R,L$ labels the  leads, $p$ denotes momentum and $l=R,L$ refers to the right and left lead, see Fig.~\ref{fig:dotsetup}(a). A current-carrying steady state is created by $\mu_L-\mu_R=eV\neq 0$. We will only consider $T_L=T_R=T$.
At $T=0$ and $V=0$, a local quantum critical point (C) at critical coupling strength separates a Kondo screening phase, governed  by a critical fixed point (MCK) from the weak coupling local moment phase (LM), see Fig.~\ref{fig:dotsetup}(b).
\begin{figure}[t!]
\centering{}\includegraphics[width=0.75\linewidth]{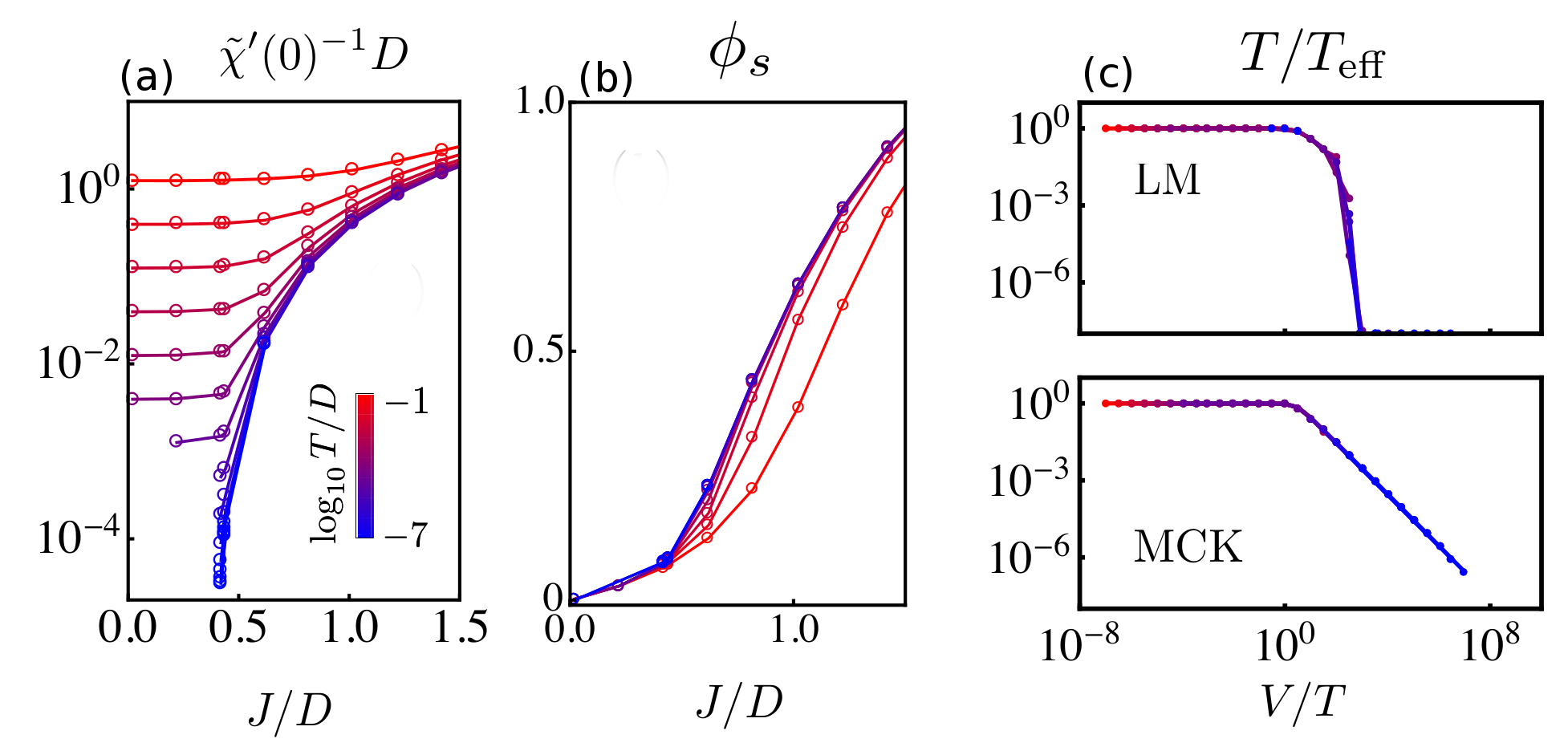}
\protect\caption{\label{fig:largeNequilibrium}
{\bf Equilibrium results and scaling of \teff at the MCK and LM fixed points}
The temperature range covers 8 decades and is encoded in the color of the curves following the color code of the inset of figure (a). The results shown here are for $r=0.15$ 
and $M/N=0.16$.
(a) Inverse of the static spin susceptibility
as a function of $J$ for different temperatures. 
(b) Singlet-strength as a function of $J$ plotted for different
values of the temperature. The inset shows that, as temperature decreases,
the $T=0$ curve is approached from below in the MCK and from above
in the LM phases.
(c)  Scaling $T_{\mbox{\tiny eff}}/T$ vs $V/T$: intermediate (MCK) and weak coupling fixed point (LM) show $T_{\mbox{\tiny eff}}\sim V$ for $V\gg T$. 
}
\end{figure}
At the QCP, Kondo screening is critically destroyed: As the QCP is approached from the MCK side, $\lim_{T \rightarrow 0}\chi^{-1}(\omega=0)\rightarrow 0$, indicative of a terminating energy scale as the local QCP is approached from the Kondo side, see Fig.~\ref{fig:largeNequilibrium}(a). This is reminiscent of what happens in several quantum critical heavy fermion compounds~\cite{Si2001,Si.14}.
In Fig.~\ref{fig:largeNequilibrium}(b) we show results for the Kondo singlet energy $\phi_{s}$~\footnote{The ground state near the MCK within the dynamical large-N is that of an overscreened Kondo effect.}~. This quantity is defined through the Kondo term contribution to the total energy, {\it i.e.} by
\begin{equation}
\frac{1}{N}\sum_{ll'}\sum_{c}J_{ll'}\av{\mathbf{S}.\mathbf{s}_{c,ll'}}=-J\kappa\left(\frac{N^{2}-1}{N}\right)\phi_{s}.
\end{equation}
This quantity requires the calculation of a higher-order correlation function. In our caseat the level of the Gaussian or mean field approximation a four-point correlator. As indicated in Fig.~\ref{fig:largeNequilibrium}(b), $\phi_s$  remains smooth through the transition and continuous in the zero-temperature limit.
 
\begin{figure}[h!]
\centering{}\includegraphics[width=1\linewidth]{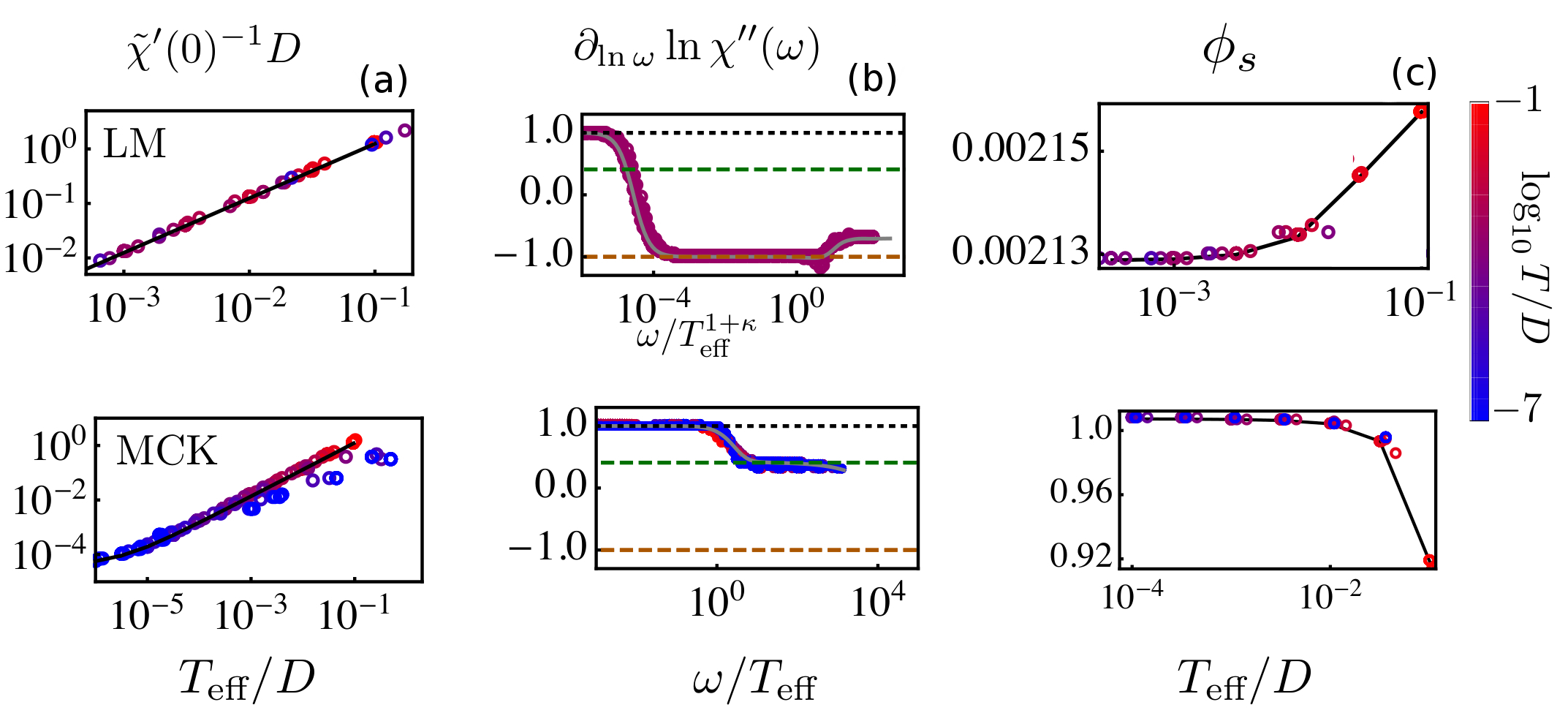}
\protect\caption{\label{fig:LargeNeffectivT}
{\bf \teff recovers the equilibrium scaling in various quantities.}
The temperature range covers 8 decades and is encoded in the color of the curves following the color code of the inset of figure (a). The results shown here are for $r=0.15$ 
and $M/N=0.16$.
Scaling of different observables with \teff for the different fixed points:
(a) Inverse static susceptibility $\chi'\left(0\right)^{-1}$vs
\teff ; (b) $\protect\partial_{\ln\omega}\ln\chi''\left(\omega\right)$
vs $\omega/T_{\mbox{\tiny eff}}$; (c) singlet strength $\phi_{s}$ vs \teff.
For each fixed point, the equilibrium scaling form (black dashed lines)
is compared with the same quantity under non-equilibrium conditions
where $T$ is substituted by \teff.
}
\end{figure}

If a \teff exists in the scaling regime of any of the fixed points (C), (MCK), (LM) of the pseudogap model, one expects that \teff $\rightarrow T$ as the nonequilibrium drive vanishes, {\it i.e.} $V\rightarrow 0$,  and all scaling functions should approach their equilibrium scaling form. On the other hand, based on arguments reminiscent to those presented above for the SDW QCP, one is led to expect
that in the large-$V$ limit \teff $\sim V$.  A frequency-dependent
\textquotedblleft effective temperature\textquotedblright , $1/\beta_{\mbox{\tiny eff}}^{A}\left(\omega\right)$,
for the observable $A$ can be defined by requiring that 
$\tanh\left[\beta_{\mbox{\tiny eff}}^{A}\left(\omega\right)\omega/2\right]=\mbox{FDR}_{A}(\omega)$~\cite{Cugliandolo.97}.
We define $T_{\mbox{\tiny eff}}$ via  the  $\mbox{FDR}_{\chi}(\omega)$ of the spin susceptibility $\chi(\omega,T)$
through
its asymptotic low-frequency behavior, $T_{\mbox{\tiny eff}}^{-1}=\lim_{\omega\to0}\beta_{\mbox{\tiny eff}}^{\chi}\left(\omega\right)$. 
It was shown in Ref.~\cite{Ribeiro.15} that \teff defined in this way indeed shows the expected behavior. Furthermore, the current-carrying steady-state  dynamic spin susceptibility scales onto the equilibrium spin susceptibility if expressed in terms of \teff, {\it i.e.}, $\Phi(\omega/T_{\mbox{\tiny eff}})$ is the steady state scaling function of $\chi(\omega,T,V)$ provided that $\Phi(\omega/T_{})$ is the equilibrium scaling function of $\chi(\omega,T)$. Here, we present data for $r=0.15$ and $M/N=0.16$. Fig.~\ref{fig:largeNequilibrium}(c) shows the scaling collapse of \teff in the variable $V/T$ at the MCK and LM. Fig.~\ref{fig:LargeNeffectivT} shows that the static and dynamic spin susceptibility as well as the Kondo singlet strength of the current-carrying steady-state at the MCK and LM fixed point assume their equilibrium scaling form when expressed  in terms of \teff. The same holds  at the Kondo-destroying QCP~\cite{Ribeiro.15}. 
It was shown in Ref.~\cite{Ribeiro.15} that even the non-linear conductance can be scaled onto the linear-response conductance when expressed in terms of \teff of the respective fixed point.

Thus, based on the dynamical large-N limit on the Keldysh contour, we conclude that an \teff exists for purely dissipative dynamics in the scaling regime near the LQC fixed point~\cite{Ribeiro.15}. This \teff is reflected in several observables. So far, we have studied the static and dynamic spin susceptibility, the steady-state conductance, and the Kondo singlet energy, a 4-point correlator within the dynamical large-N approach. In each case, replacing $T$ in the equilibrium scaling function by \teff results in the steady-state scaling function of that quantity~\cite{Ribeiro.15}. We stress that both the QCP (C) and the (MCK) are intermediate coupling fixed points that are fully interacting within our approach. Thus, we expect that $1/N$ corrections will not modify our conclusions. This is in line with the observation that our approach predicts the correct equilibrium scaling function of the $N=2$ pseudogap model~\cite{Glossop.11}.

A particularly observation is that even near the weak-coupling LM fixed point
 \teff is an observable independent quantity that reproduces the equilibrium scaling functions.  This fixed point shares features of the SDW QCP and the classical fixed point analyzed by Calabrese and Gambassi~\cite{Calabrese.04} as any residual interaction vanishes as the fixed point is approached. To analyze this situation further, we investigate the existence of \teff in the non-interacting pseudogap resonant level model (RLM). 

\section{The pseudogap resonant level model}
\label{Sec:RLM}
\begin{figure}[h!]
\centering{}\includegraphics[width=1\linewidth]{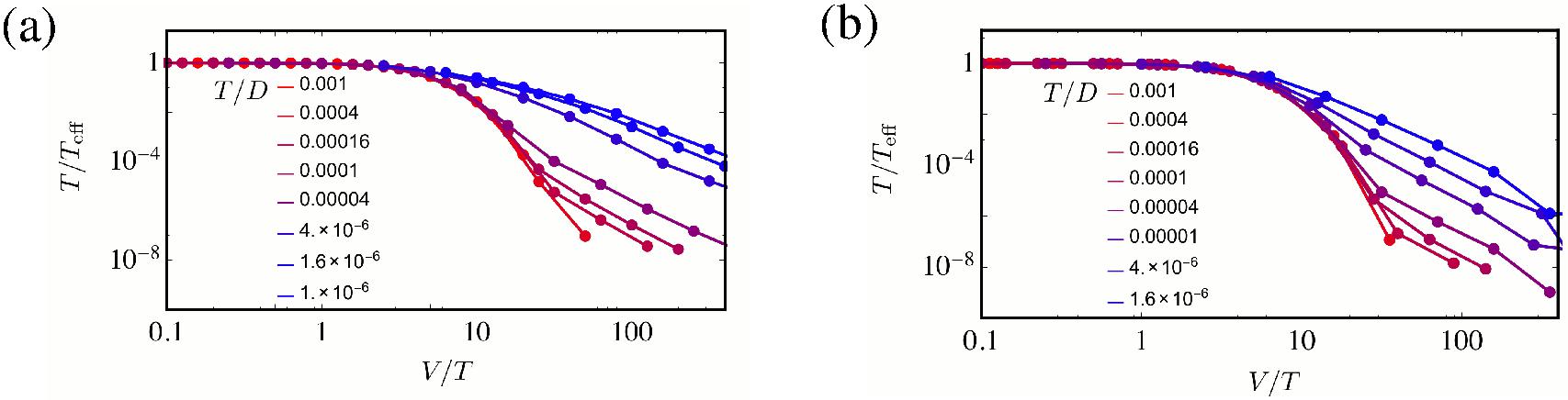}
\protect\caption{\label{fig:RLM}
{\bf \teff for the pseudogap resonant level model.}
(a) Scaling plot of \teff for the pseudogap RLM, Eq.~(\ref{Eq:RLM}) with $\epsilon_d=0$, $D=10$, $\tilde{V}_L=\tilde{V}_R=0.1$ and powerlaw exponent $r=0.2$. Scaling collapse is only found for
\teff$=T$, {\it i.e.} in the linear response regime. Away from linear response, $T/$\teff does not collapse as a function of $V/T$.
(b)  same as in (a) expect for a  powerlaw exponent $r=0.3$.
}
\end{figure}

The resonant level model (RLM) often serves as a proxy for the fully interacting Anderson impurity model. The an-isotropic Kondo model at its Toulouse point can be mapped onto the (spinless) RLM. A mean-field analysis of the pseudogap Anderson model in the slave boson representation leads to the pseudogap RLM augmented with a selfconsistency condition. The Hamiltonian of the model is
\begin{equation}
\label{Eq:RLM}
H_{\mbox{\tiny RLM}}=\sum_{k,\sigma}\varepsilon_k c^\dagger_{k,\sigma} c^{}_{k,\sigma}+ \sum_{\sigma} \varepsilon_d d^\dagger_\sigma d^{}_\sigma +\sum_{k,\sigma} \tilde{V}_\alpha \Big ( c^{\dagger}_{k,\sigma} d^{}_\sigma + d^\dagger_\sigma c^{}_{k,\sigma} \Big).
\end{equation}
We will assume particle-hole symmetry and that the density of states of the conduction electrons vanishes in a powerlaw fashion at the Fermi energy ($\omega=0$), $\rho(\omega)=\sum_{\epsilon_k}\delta(\epsilon_k-\omega)\sim |\omega|^r\Theta(D-|\omega|)$ with powerlaw exponent $0<r<1$. This corresponds to $U=0$ and $\epsilon_d=0$ in the sketch of Fig.~\ref{fig:dotsetup}(a).
In equilibrium this leads to  powerlaw behavior in the $\omega$-behavior of the T-matrix $|V|^2 G^{R}(\omega,T)$ and dynamic spin susceptibility $\chi(\omega,T)$ as $\omega \rightarrow 0$. From Eq.~(\ref{Eq:RLM}), we find
\begin{eqnarray}
\label{Eq:chiKeldyshRLM}
\chi^{K}(\omega,T,V)\!\!\!\!\!\!&=&\!\!\!\!\!\!  \int d\epsilon \Bigg [ \frac{\sum_\alpha \Gamma^\alpha(\epsilon)f(\epsilon-\mu_\alpha)}{[\epsilon-\Lambda(\epsilon)]^2+[\Gamma(\epsilon)]^2}
\times \frac{\sum_\alpha \Gamma^\alpha(\epsilon-\omega)f(\omega-\epsilon+\mu_\alpha)}{[\epsilon-\omega-\Lambda(\epsilon-\omega)]^2+[\Gamma(\epsilon-\omega)]^2} \\
\!\!\!&+&\!\!\! \frac{\sum_\alpha \Gamma^\alpha(\epsilon-\omega)f(\epsilon-\omega-\mu_\alpha)}{[\epsilon-\Lambda(\epsilon)]^2+[\Gamma(\epsilon)]^2}
\times \frac{\sum_\alpha \Gamma^\alpha(\epsilon)f(-\epsilon+\mu_\alpha)}{[\epsilon-\omega-\Lambda(\epsilon-\omega)]^2+[\Gamma(\epsilon-\omega)]^2} \Bigg ],\nonumber 
\end{eqnarray}
where $\Gamma(\epsilon)=\sum_{\alpha=L,R}\Gamma^\alpha(\epsilon)=\pi \sum_{\alpha=L,R}|\tilde{V}_\alpha|^2_\alpha \sum_k \delta(\epsilon_k-\epsilon)$ and $\Lambda(\epsilon)$ is the Hilbert transform of $\Gamma(\epsilon)$, $f(x)$ is the Fermi function and $eV=\mu_L-\mu_R$. 
Eq.~(\ref{Eq:chiKeldyshRLM}) allows us to analyze and compare $(\chi^{-1})^{K}$ with the results of Ref.~\cite{Mitra.06}, {\it i.e.} with Eq.~(\ref{ChiKeldysh}). We find that 
in the large-$V$ limit $(\chi^{-1})^{K}$ approaches a $V$ dependent constant which thus permits us to introduce an \teff in complete analogy to the procedure that led to Eq.~(\ref{Eq:TEFF}). This requires the construction of
$\chi^R(\omega,T,V)$ and $\chi^A(\omega,T,V)$ and the associated FDR. Deviations from scaling occur for the pseudogap Kondo model for temperature or frequency of the scale of the Kondo temperature $T_K$ of the $r=0$ Kondo model. Thus,  the scaling regime is obtained for $T,\omega\ll T_K$. The resonant level width $\Gamma$ of the RLM corresponds to $T_K$ of the Kondo model. We thus will only consider $T<\Gamma$ and $V<\Gamma$.
The resulting $T/$\teff is shown in Fig.~\ref{fig:RLM} as a function of $V/T$ for two different powerlaw exponents; $r=0.2$ in Fig.~\ref{fig:RLM}(a) and $r=0.3$ in Fig.~\ref{fig:RLM}(b). Our results suggest that a scaling collapse reminiscent of what is found in the fully interacting pseudogap Kondo model is absent~\cite{Ribeiro.15}.

\section{Conclusion}
We have reviewed the non-thermal steady state dynamics of two models of quantum criticality, the spin-density wave theory and the pseudogap Kondo impurity model, a model of local quantum criticality. In both cases, voltage and temperature are relevant perturbations and in both cases only the  case of simple relaxational dynamics was considered.  An analysis of the relation between correlation and response functions in terms of   effective termperatures, valid in the scaling regime accompanying these quantum critical points,  has been presented. For the SDW, an effective temperature exists for the order parameter susceptibility in the limit of large bias voltage. For the pseudogap Kondo model we find that several scaling functions, including that of a four-point correlator, reproduce their equilibrium behavior when rescaled in terms of a uniquely defined \teff. This \teff has the additional property that it shows scaling collapse of $T/$\teff in terms of $V/T$. 
It is important to note that the results for the two models are based on different methodologies. The SDW results were obtained from a perturbative RG scaling approach on the Keldysh contour~\cite{Mitra.06,Mitra.08} while the results for the critical Kondo destruction were obtained within a dynamical large-N limit~\cite{kirchner2010,Ribeiro2013b,Ribeiro.15}
Interestingly, there are indications that the fluctuation spectrum of current-carrying steady states is that of a thermal state within the ADS/CFT correspondence~\cite{Sonner.12,Bhaseen.15}.
We complemented our analysis by results for \teff obtained for the pseudogap resonant level model, an non-interacting relative of the pseudogap Kondo model. 
The pseudogap resonant level model is a quantum impurity model that emerges e.g. in a mean-field treatment of the pseudogap Kondo model. As a consequence of its non-interacting nature its steady-state properties are derived from an action of the form of Eq.~\ref{Eq:GaussianKeldysh}. Although  we find that $(\chi^{-1})^{K}$ of the pseudogap resonant level model approaches a $V$ dependent constant so that  a \teff in the large-V limit can be defined a scaling collapse of $T/$\teff in terms of $V/T$ was not found for $T<\Gamma$ and $V<\Gamma$. This suggests that the behavior near the Gaussian fixed point differs from that obtained near the critical fixed points of the pseudogap Kondo model reported in Ref.~\cite{Ribeiro.15}.

\noindent
\paragraph{Acknowledgments}
We thank Q.~Si  for useful discussions. P.~Ribeiro acknowledges support by FCT through the Investigador FCT contract IF/00347/2014.
S.~Kirchner acknowledges partial support by the National Science Foundation of China, grant No.11474250.

\section*{References}

\end{document}